\documentclass[aps,prl,twocolumn,superscriptaddress]{revtex4}
\usepackage{epsfig}

\begin{document}
\title{Crossover from 2D to 3D magnetic disorder in sub-mono-atomic
ferromagnetic layers}
\author{A. Frydman}
\address{Department of Physics, Bar Ilan University, Ramat Gan 52900,
Israel}
\author{R.C. Dynes}
\address{Department of Physics , University of California, San Diego, La Jolla, CA
92093}

\begin{abstract}
We present transport and magnetoresistance (MR) measurements
performed on quench condensed ultrathin films of Gd evaporated on
an amorphous Ge or Sb layer. These films show a large negative MR
accompanied by a hysteretic superimposed structure. When the film
is coated by an overlayer of Ge or Sb the magnitude of the MR
increases and the hysteretic structure disappears. We speculate
that the findings are a result of a crossover from a 2D magnetic
disorder in the uncoated layers of Gd to a 3D magnetic disorder as
the Gd film is coated by a semiconducting layer.
\end{abstract}
\pacs{75.70.Ak,75.50.Pp,71.23.Cq}
 \maketitle

The magneto-transport properties of magnetic rare-earth atoms
imbedded in a non-magnetic insulating matrix is a dramatic example
of the effect of magnetic moments on the conductivity of
semiconductors. Depending on the magnetic atom concentration,
these systems can show extremely large magnetoresistance values
\cite{von1,von2,washburn}. In the case of three dimensional
amorphous Gd-Si where the Gd ion has a spin s=7/2 it has been
shown that the system can be continuously tuned through the
metal-insulator transition \cite{Teizer1}. The observed magnetic
field driven transition from localized states subject to variable
range hopping to extended states which show metallic behavior, was
interpreted as being due to the extra degree of freedom added by
the spin on the Gd. In the absence of magnetic field the Gd spin
is randomly oriented and thus adds an additional disorder not
experienced by a non-spin system \cite{ieee}. With increasing
magnetic field, the Gd spins are driven towards alignment,
reducing the net disorder and thus increasing the conductivity. On
the insulating side of the metal insulator transition, the
magnetoresistance is spectacularly exponentially dependent on
temperature
and exceeds 5 orders of magnitude for a 10T field at T%
\mbox{$<$}%
100mK \cite{peng}.

In addition, tunneling and hall measurements were performed on these systems %
\cite{Teizer2}. They show that the density of states (DOS) is
substantially modified by the presence of magnetic field. For an
insulating Gd-Si alloy in low field there are very few, if any,
states available to tunnel into at the Fermi level. The tunneling
conductance exhibits a strong voltage (energy) dependence
demonstrating that this is a strongly coulomb correlated system.
As the field increases, the tunneling conductance increases such
that a very clear correlation between conductivity and density of
states is found. Hall measurements show a striking field
dependence of the hall coefficient with 1/Rh collapsing to zero at
the metal insulator transition as well.

A number of experimental findings are noteworthy. First, the
negative MR observed in these systems continues to 20T without
showing signs of
saturation. This is especially surprising since Gd is a trivalent 4{\it f}$%
^{7}$5{\it d}$^{1}$6{\it s}$^{2}$ atom which is characterized by
J=S=7/2 and L=0 due to the half filled f shell. Hence, no magnetic
anisotropy is expected for the Gd ions and the magnetoresistance
is expected to saturate at a relatively low magnetic field.
Secondly, magnetization measurement
show that the moment in these alloys is usually smaller than 7/2 \cite%
{hellman1}. It peaks at the metal insulator transition (MIT) where
it approaches the bulk value of 7/2 but on both sides of the
transition it is substantially reduced. These two findings lead to
the speculation that there is an antiferromagnetic coupling
between the Gd atoms and the surrounding semiconducting atoms
\cite{hellman2} leading both to the lack of high field saturation
and to the reduced net moment. The peak at the MIT is still not
understood.

\begin{figure}\centering
\epsfxsize12cm\epsfbox{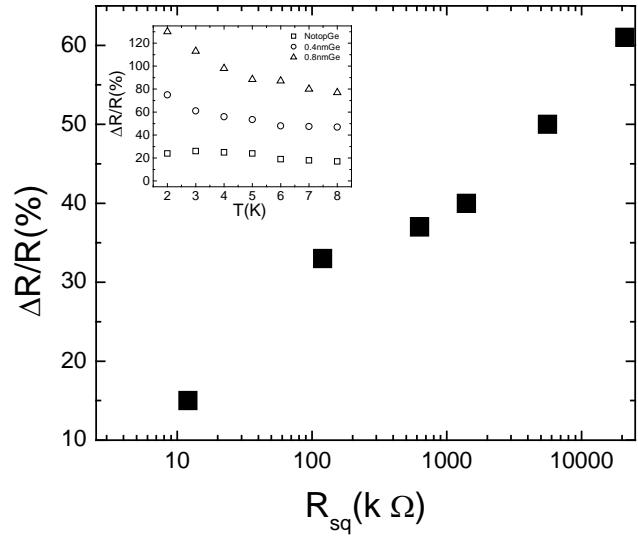} \vskip 0truecm\caption
{Magnetoresistance magnitude ($\frac{R_{0}-R_{9T}}{R_{9T}})$ as a
function of sheet resistance for an uncoated Gd sub monolayer at
T=4K (In this sample the maximum resistance was attained at H=0).
The insert shows the temperature dependence of $\Delta R/R$ in a
different sample for the uncoated Gd (squares, R$_{sq}(4K)$=0.5
M$\Omega ),$an overlayer of 4\AA\ thick Ge (circles,
R$_{sq}(4K)$=1.2 M$\Omega )$and for an overlayer of 8\AA\
(triangles, R$_{sq}(4K)$=0.6 M$\Omega ).$Note that the temperature
dependence becomes more pronounced with overlayer thickness.
 }\end{figure}

Very little has been explored in two dimensions along these lines.
In this paper we describe an experiment performed on a sub
mono-atomic layer of Gd grown on an amorphous Ge or Sb substrate.
We present MR data of this film. In addition, we study the change
in the MR as we subsequently deposit an overlayer of amorphous Ge
or Sb, {\it in-situ}. We find that there are large differences
between an uncoated Gd sub monolayer system and one which is
coated by an overlayer. The results demonstrate the important role
played by the semiconductor surroundings on the magnetic
properties of the ferromagnetic-semiconductor system.

 The samples described in this paper were prepared by quench
condensation i.e. sequential evaporation on a cryogenically cold
substrate under UHV conditions while monitoring the film thickness
and resistance. It is well established that if a metallic material
is quench-condensed on a substrate that is pre-coated by an
ultra-thin buffer layer of amorphous Ge or Sb, the sample grows
rather uniformly and is electrically continuous at a thickness of
1-2 monolayers of material\cite{strongin}. This is in contrast
with the growth without a buffer layer in which case the film
grows in a granular morphology \cite{granular bob,granular
goldman,granular rich} The quench condensation method provides a
very sensitive control on the sample growth process, allowing one
to terminate the evaporation at any desired stage of the material
deposition and ''freeze'' the morphological configuration. In
particular, this technique allows one to stop the film growth at a
thickness for which a measurable conductivity first appears across
the sample. Hence one can grow an ultrathin layer which is
electrically continuous but geometrically it is still a
sub-monolayer. At T=0 we expect such monolayers to be insulating.
In the current work we prepared the samples by growing a 10\AA\
thick Ge or Sb buffer-layer on a Si/SiO substrate at T=5K. Then we
quench condensed an
ultrathin layer of Gd (3-4 \AA\ thick) until a sheet resistance of a few M$%
\Omega $ was measured across the sample. We then added sequential
sub monolayers of Gd thus reducing the sheet resistance, R$_{sq}$,
in a controlled fashion. Finally, we coated the Gd by overlayers
of the amorphous semiconductor material (Ge or Sb). Throughout the
experiment we measured the transport (R-T) and magnetoresistance
(R-H) properties for each evaporated layer. Magnetic field was
applied perpendicular to the film. All measurements were performed
in a shielded room and we ensured that the I-V characteristics of
all films were in the ohmic regime.

\begin{figure}\centering
\epsfxsize11cm\epsfbox{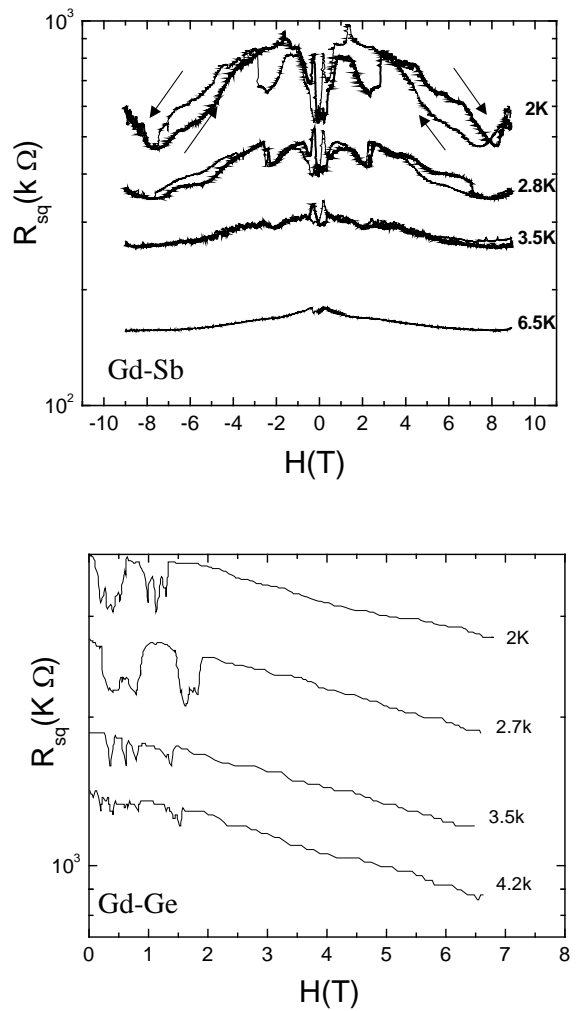} \vskip -1truecm\caption
{Magnetoresistance traces of uncoated Gd sub monolayers on
amorphous Sb (top) and Ge (bottom) for different temperatures. The
top frame shows MR for both magnetic field sweep directions. Solid
lines are for sweeps from left to right and dotted lines are for
sweeps from right to left. The hysteretic nature of the traces is
obvious.
 }\end{figure}

We have studied 9 sets of samples of Gd-Sb and Gd-Ge, all of these
samples exhibit similar behavior. The uncoated Gd sub-atomic
layers on a semiconductor substrate exhibit negative MR which
persist to 9 tesla without saturation. We observe amplitudes of
the MR ($\frac{R_{0}-R_{9T}}{R_{9T}})$ up to a factor of 60\% at
T=2K, decreasing monotonically as Gd is added to the layer, thus
reducing R$_{sq}$ (see figure 1). This is similar to the behavior
in the 3D alloys in which the MR magnitudes decreases as the Gd
concentration is increased. However, the 2D samples show a number
of unique properties. The temperature dependence of the MR in the
uncoated Gd layer at H=9T often tends to saturate as the
temperature is reduced below T=4K as demonstrated in figures 1 and
2 (left panel). In addition, a sample dependent structure is
superimposed on the usual smooth negative MR background. This can
be seen in figure 2 which shows a very profound detailed and
hysteretic MR trace. The curves trace each other when sweeping the
field in the same direction many times but they produce hysteretic
mirror images when sweeping the field in opposite directions. The
structure persists, in many cases, up to a field of 9T (the
highest available field in our experiment). This is very
surprising since a hysteretic MR trace is associated with magnetic
anisotropy and preferred magnetic moment orientation. Since Gd is
believed to be spherical, any anisotropy is expected to be
negligible and its effect is expected to be limited to very small
magnetic fields. This is clearly not the case for the results
shown in figure 2 where the hysteretic sharp structure persists to
at least several tesla and to at least 4.2K. None of this
hysteresis is observed in 3 dimensions.

As the Gd layer is subsequently coated by an overlayer of Ge or Sb
a number of clear trends are observed. The initial layers of the
semiconductor overlayer (up to a thickness of $\sim 4$\AA ) cause
both the resistance and the MR to increase. Additional Ge or Sb
overlayer evaporation causes the resistance to then decrease,
however the MR continues to increase (we have observed
$\frac{\Delta R}{R}$ values up to a factor of 5 at 9T) until, for
thick enough overlayers (thicker than approximately 10\AA ), $\frac{\Delta R%
}{R}$ does not change with additional overlayer. In addition, the
overlayers of Ge or Sb suppress the hysteretic structure and
eventually yield smooth negative MR curves as demonstrated in
figure 3. Another striking effect of the semiconductor overlayer
is associated with the temperature dependence of the resistance.
Figure 4 shows the transport properties of a Gd-Ge sample for
uncoated Gd and two overlayer thicknesses for H=0 and H=9T. It is
seen that the R-T curves can loosely fit a $\sigma =c\cdot T$
dependence in this temperature range. The slope, c, changes as a
function of the overlayer thickness so that the R-T curves cross
each other at various temperatures. This is very different than
the behavior in the 3D case where the temperature dependence
becomes weaker (monotonically) as the resistance decreases. In the
2D case, the observed crossing of the R-T traces implies that the
overlayer material modifies the conduction mechanism leading to a
qualitatively different temperature dependence.

 \begin{figure}\centering
\epsfxsize11cm\epsfbox{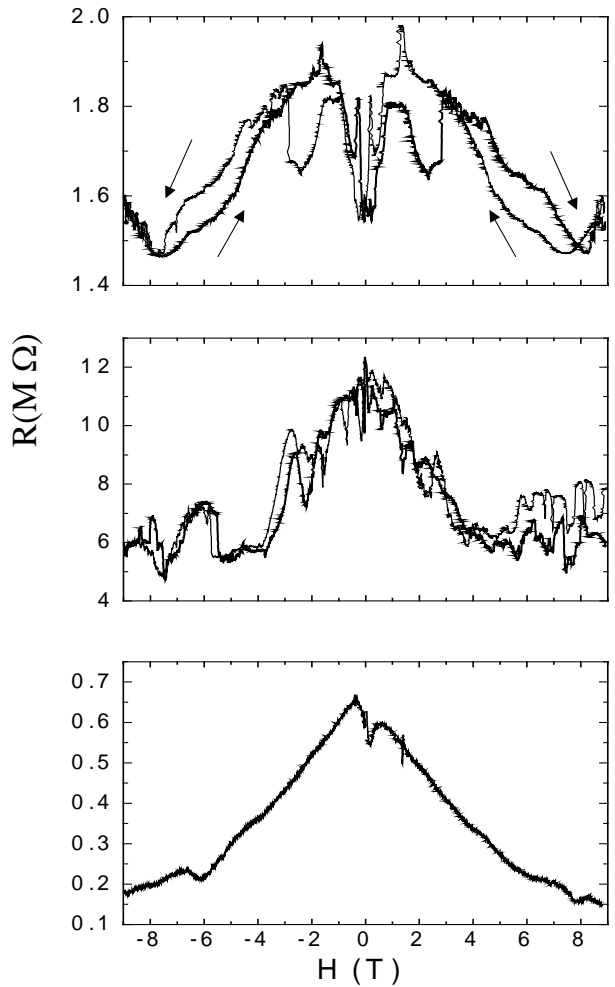} \vskip -1truecm\caption
{Magnetoresistance curves of the Gd-Sb sample of figure 2 for the
uncoated layer (top), a 5\AA\ thick Sb overlayer (middle) and a
9\AA\ thick Sb overlayer (bottom).}\end{figure}

The non-monotonic dependence of the resistance with overlayer
thickness is peculiar. Naively, all other thing being unchanged,
one would expect that adding a semiconductor between the Gd atoms
would decrease the resistance because of the decrease in the
tunneling barrier heights between the metallic atoms. This is
indeed observed for thicker overlayers. The fact that the
resistance increases for initial overlayer thicknesses implies
that there is another competing mechanism. The observation that
adding the overlayer is accompanied by a significant increase in
MR amplitudes (for both regions of resistance change) leads us to
speculate that the increase of resistance in the first overlayer
stages is associated with an increase in magnetic disorder.We
envision that the uncovered Gd atoms are forced to be oriented
perpendicular to the substrate. The chemical bond of the Gd atom
to the substrate semiconductor breaks the natural magnetic
symmetry of the ion and orients the moment perpendicular to the
substrate. Ferromagnetic interactions between the Gd atoms create
magnetic domains which are either ''up'' or ''down'' on the
surface. In this picture, aside from the up/down disorder, much of
the magnetic disorder, which characterizes the 3D alloys and
dominates their MR behavior, is quenched in the 2D sub-mono-atomic
films because the Gd atoms moments are confined to being vertical.
The subsequent semiconductor overlayer relaxes this constraint and
allows the magnetic moment to splay into all the 3D space. This
increases the magnetic disorder leading to an increase both in the
resistance and in the MR. It appears therefore that there are two
conflicting effects of the top semiconducting layer. The
generation of parallel tunneling paths acts to raise the
conductivity and the increase of magnetic disorder acts to reduce
conductivity. The latter effect is highly dependent on temperature
as observed in the 3D alloy case, while tunneling effects are
relatively temperature insensitive. This may explain the unique
temperature dependence of the resistance as an overlayer is added
to the Gd film (figure 4). At high temperatures the main effect of
the semiconducting layer is to increase the tunneling conductance,
while at low temperatures the dominant factor is the enhancement
of disorder which decreases conductivity. This leads to the
crossing of the R-T curves as a function of overlayer thickness
which is observed in figure 4.

\begin{figure}\centering
\epsfxsize8.5cm\epsfbox{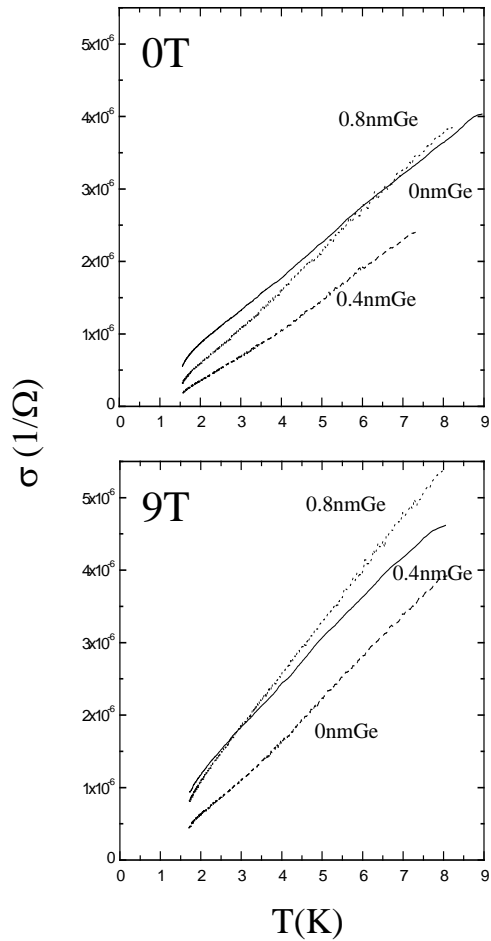} \vskip 7truecm\caption
{Conductivity as a function of temperature for a Gd-Ge sample at
H=0T (top panel) and H=9T (bottom panel). Solid lines are for the
uncoated Gd layer, dashed lines are for a 4\AA\ thick Ge overlayer
and dotted lines are for a 8\AA\ thick Ge overlayer.
 }\end{figure}

The above picture can also explain the evolution of the structure
observed in the MR (figure 3). The hysteretic structure is a
consequence of the anisotropy in the Gd atoms on the surface. In
the absence of magnetic field, domains are randomly oriented up or
down. With the application of a perpendicular field, the domains
orient with the field direction. Each flip of a domain moment will
cause a sharp resistance change (which may be an increase or
decrease depending on the microscopic configuration). The presence
of an overlayer relaxes the symmetry and restores the isotropic
nature of the Gd. In effect, coating the Gd with amorphous Ge or
Sb causes a crossover from a 2D magnetic disorder to a 3D magnetic
disorder. This leads to suppression of the hysteretic structure
because in the 3D disorder the moments can rotate continuously,
without abrupt orientation flips, until they align with the field.

In summary, the magnetic properties of a 2D sub monolayer of Gd on
an amorphous semiconducting substrate show a number of features
which are different than the 3D alloys. These include a relatively
small magnitude of MR, small temperature dependence and a
hysteretic fine-structure superimposed on the MR curves. When
these layers are covered by a semiconducting overlayer, the
samples evolve towards behavior which is more characteristic of 3D
alloys, i.e. the MR and its temperature dependence becomes larger
and the fine-structure vanishes. These results lead us to suggest
that the coated layers are the limiting case of a very thin Gd-Ge
alloy and share similar physics while the uncoated layers exhibit
qualitatively different results due to the fact that the magnetic
disorder in these samples is substantially reduced.

We gratefully acknowledge illuminating discussions with F.
Hellman. This research was supported by the Binational USA-Israel
fund grant number 1999332 and the NSF grant DMR0097242.

\end{document}